\documentstyle[12pt]{article}
\newcommand{\beq}{\begin{equation}}
\newcommand{\eeq}{\end{equation}}
\newcommand{\beqa}{\begin{eqnarray}}
\newcommand{\eeqa}{\end{eqnarray}}
\newcommand{\ba}{\begin{array}}
\newcommand{\ea}{\end{array}}
\newcommand{\CR}{\nonumber \\}
\newcommand{\pa}{\partial}

\newcommand{\D}{\delta}

\newcommand{\La}{\Lambda}

\newcommand{\bra}{\langle}
\newcommand{\ket}{\rangle}
\newcommand{\lm}{\lambda}

\newcommand{\tQ}{\tilde{Q}}
\newcommand{\rep}{{\cal R}}

\begin{document}

\begin{titlepage}
\null
\begin{flushright} 
9803014  \\
UTHEP-364  \\
March, 1998
\end{flushright}
\vspace{0.5cm} 
\begin{center}
{\Large \bf
Exceptional Seiberg-Witten Geometry with \\ Massive Fundamental Matters
\par}
\lineskip .75em
\vskip2.5cm
\normalsize
{\large Seiji Terashima and Sung-Kil Yang} 
\vskip 1.5em
{\large \it Institute of Physics, University of Tsukuba, 
Ibaraki 305-0006, Japan}
\vskip3cm
{\bf Abstract}
\end{center} \par
We propose Seiberg-Witten geometry for $N=2$ gauge theory with gauge 
group $E_6$ with massive $N_f$ fundamental hypermultiplets. 
The relevant manifold is described as a fibration of the ALE space of $E_6$ 
type. It is observed that the fibering data over the base ${\bf CP}^1$ has an
intricate dependence on hypermultiplet bare masses.
\end{titlepage}

\baselineskip=0.7cm

Recently several attempts have been made in extending the work of
Seiberg and Witten  on four-dimensional $N=2$ supersymmetric gauge theory 
\cite{SeWi},\cite{review} so as to include matter hypermultiplets in
representations other than the fundamentals \cite{TeYa2}-\cite{OdToSaSa}.
In our previous paper \cite{TeYa2} we have refined the technique of 
$N=1$ confining phase superpotentials toward the application to $N=2$ or $N=1$
gauge theories with A-D-E gauge groups with matter hypermultiplets in addition
to an adjoint matter. The moduli space of $N=1$ confining phase can also be
studied in view of $M$-theory fivebrane, the result of which is in
agreement with the field theory analysis \cite{DeOz},\cite{Te}.
In \cite{LL} the fivebrane configurations are considered 
to obtain complex curves describing $N=2$ $SU(N_c)$ gauge theory with 
matters in the symmetric or antisymmetric representation. 
In \cite{AG} the approach based on Type IIA string
compactification on Calabi-Yau threefolds is used to find exact descriptions
of $N=2$ $SO(N_c)$ gauge theory with vectors and spinors.

In this paper we propose Seiberg-Witten geometry 
for $N=2$ supersymmetric gauge theory with gauge group $E_6$ 
with massive $N_f$ fundamental hypermultiplets. To this aim we employ the 
technique of $N=1$ confining phase superpotentials \cite{EFGIR}-\cite{GPR}. 
The ALE space description of Seiberg-Witten geometry for $N=2$ $SU(N_c)$
and $SO(2N_c)$ QCD is recovered in this approach \cite{TeYa2}.  We will show 
in the following that an extension of \cite{TeYa2} enables us to obtain 
exceptional Seiberg-Witten geometry with fundamental hypermultiplets.
The resulting manifold  takes the form of a fibration of the ALE space of 
type $E_6$.

Let us consider $N=1$ $E_6$ gauge theory with $N_f$ fundamental matters
$Q^i,\tQ_j$ ($1 \leq i,j \leq N_f$) and an adjoint matter $\Phi$. $Q^i,\tQ_j$
are in ${\bf 27}$ and $\overline {\bf 27}$, and $\Phi$ in ${\bf 78}$ of $E_6$.
The coefficient of the one-loop beta function is given by $b=24-6N_f$, and 
hence the theory is asymptotically free for $N_f=0, 1, 2, 3$ and finite
for $N_f=4$. We take a tree-level superpotential
\beq
W = \sum_{k \in {\cal S}} g_{k} s_{k}(\Phi) +
{\rm Tr}_{N_f} \, \gamma_0 \, \tilde{Q} Q +
{\rm Tr}_{N_f} \, \gamma_1 \, \tilde{Q} \Phi Q,
\label{treem}
\eeq
where ${\cal S}=\{2,5,6,8,9,12\}$ denotes the set of degrees of 
$E_6$ Casimirs $s_k(\Phi)$ and
$g_k$, $(\gamma_a)_i^j$ $(1 \leq i,j \leq N_f)$ are coupling 
constants. A basis for the $E_6$ Casimirs will be specified momentarily.
When we put $(\gamma_0)^i_j=\sqrt{2} m^i_j$ with $[m, m^{\dagger}]=0$,
$(\gamma_1)^j_i=\sqrt{2} \D^j_i$ and all $g_k=0$, (\ref{treem}) is reduced to 
the superpotential in $N=2$ supersymmetric Yang-Mills theory with massive
$N_f$ hypermultiplets.

We now look at the Coulomb phase with $Q=\tilde{Q}=0$.
Since $\Phi$ is restricted to take the values in the Cartan subalgebra we
express the classical value of $\Phi$ in terms of a vector \footnote{Our
notation is slightly different from \cite{TeYa2}. Here we use $a_i$ with
lower index instead of $a^i$ in \cite{TeYa2}.}
\beq
a=\sum_{i=1}^6 a_i\alpha_i
\eeq
with $\alpha_i$ being the simple roots of $E_6$. 
Then the classical vacuum is parametrized by
\beq
\Phi^{cl}={\rm diag}\, (a\cdot \lm_1, a\cdot \lm_2,\cdots , a\cdot \lm_{27}),
\eeq
where $\lm_i$ are the weights for ${\bf 27}$ of $E_6$. 
For the notation of roots and weights we follow \cite{Sla}. 
We define a basis for the $E_6$ Casimirs $u_k(\Phi)$ by
\beqa
&& u_2=-{1\over 12}\, \chi_2, \hskip10mm u_5=-{1\over 60}\, \chi_5,
\hskip10mm u_6=-{1\over 6}\, \chi_6+{1\over 6\cdot 12^2}\, \chi_2^3,   \CR
&& u_8=-{1\over 40}\, \chi_8+{1\over 180}\, \chi_2\chi_6
-{1\over 2\cdot 12^4}\, \chi_2^4, \hskip10mm
u_9=-{1\over 7\cdot 6^2}\, \chi_9+{1\over 20\cdot 6^3}\, \chi_2^2\chi_5, \CR
&& u_{12}=-{1\over 60}\, \chi_{12}+{1\over 5\cdot 6^3}\, \chi_6^2
+{13\over 5\cdot 12^3}\, \chi_2\chi_5^2 \CR
&& \hskip11mm +{5\over 2\cdot 12^3}\, \chi_2^2\chi_8
-{1\over 3\cdot 6^4}\, \chi_2^3\chi_6+{29\over 10\cdot 12^6}\, \chi_2^6,
\eeqa
where $\chi_n={\rm Tr}\, \Phi^n$. 
The standard basis $w_k(\Phi)$ are written in terms of $u_k$ as follows
\beqa
&& w_2={1\over 2}u_2, \hskip10mm w_5=-{1\over 4}u_5, 
\hskip10mm w_6={1\over 96}\left( u_6-u_2^3\right), \CR
&& w_8={1\over 96}\left( u_8+{1\over 4}u_2u_6-{1\over 8}u_2^4 \right),
\hskip10mm w_9=-{1\over 48} \left( u_9-{1\over 4} u_2^2u_5\right), \CR
&& w_{12}={1\over 3456} \left( u_{12}+{3\over 32}u_6^2-{3\over 4}u_2^2u_8
-{3\over 16}u_2^3 u_6+{1\over 16}u_2^6 \right).
\label{casimiro}
\eeqa
The basis $\{u_k\}$ and (\ref{casimiro}) were first introduced 
in \cite{LW}.\footnote{The Casimirs $u_1,u_2,u_3,u_4,u_5,u_6$
in \cite{LW} are denoted
here as $u_2,u_5,u_6,u_8,u_9,u_{12}$, respectively.}
In our superpotential (\ref{treem}) we then set
\beq
s_2=w_2, \;\; s_5=w_5, \;\; s_6=w_6, \;\; s_8=w_8, \;\; s_9=w_9, \;\; 
s_{12}=w_{12}-\frac{1}{4} w_6^2.
\label{casimir}
\eeq
We will discuss later why this particular form is assumed.

The equations of motion are given by
\beq
\frac{\pa W(a)}{\pa a_i}=
\sum_{k \in {\cal S}} g_{k} \frac{\pa s_{k}(a)}{\pa a_i}=0.
\label{eq1}
\eeq
Let us focus on the classical vacua with
an unbroken $SU(2) \times U(1)^{5}$ gauge symmetry. Fix the $SU(2)$ 
direction by choosing the simple root $\alpha_1$, then we have the vacuum 
condition
\beq
a\cdot \alpha_1=2a_1-a_2=0.
\label{vaccon}
\eeq
It follows from (\ref{eq1}), (\ref{vaccon}) that
\beqa
&& {g_9\over g_{12}}={D_{1,9}\over D_{1,12}}   \CR
&&  \hskip7mm = -{1\over 8} 
\Big( 2 a_1 a_5 a_4-a_4 a_3^2+a_5^2 a_4+a_4^2 a_3-a_3 a_6^2+a_3^2 a_6  \CR
&& \hskip20mm -2 a_1 a_5^2+2 a_1 a_6^2-2 a_4^2 a_1-a_5 a_4^2
-2 a_1 a_3 a_6+2 a_4 a_3 a_1  \Big), \CR
&& {g_8\over g_{12}}={D_{1,8}\over D_{1,12}}  \CR
&&  \hskip7mm = -{1\over 48}        
\Big( 12 a_1 a_5^2 a_4-6 a_1^2 a_5^2-6 a_1^2 a_6^2-4 a_1^3 a_3+4 a_3^3 a_1
+2 a_3^3 a_4+2 a_3^3 a_6-a_4^4  \CR
&& \hskip20mm -3 a_3^2 a_6^2-3 a_3^2 a_4^2-a_3^4-a_6^4
-12 a_1 a_5 a_4^2-2 a_5 a_4 a_6^2+8 a_1 a_3 a_6^2+3 a_1^4 \CR
&& \hskip20mm  +6 a_1^2 a_4 a_3-8 a_4 a_3^2 a_1-2 a_5 a_4 a_3^2-2 a_4^2 a_3 a_6
+6 a_1^2 a_5 a_4-2 a_4 a_3 a_6^2  \CR
&& \hskip20mm +2 a_5 a_4^2 a_3-2 a_5^2 a_3 a_6
+2 a_4 a_3^2 a_6-a_5^4-2 a_5^2 a_4 a_3-2 a_1^2 a_3^2-6 a_1^2 a_4^2  \CR
&& \hskip20mm +2 a_5^2 a_6^2+2 a_4^2 a_6^2+2 a_5^2 a_3^2
+6 a_1^2 a_3 a_6-8 a_1 a_3^2 a_6+8 a_4^2 a_3 a_1-4 a_5^2 a_1 a_3  \CR
&& \hskip20mm +2 a_5 a_4 a_3 a_6
+4 a_5 a_4 a_1 a_3+2 a_3 a_4^3-3 a_4^2 a_5^2
+2 a_4 a_5^3+2 a_4^3 a_5+2 a_6^3 a_3 \Big) ,   \CR
&& {g_6\over g_{12}}={D_{1,6}\over D_{1,12}}   \CR
&&  \hskip7mm = {1\over 192} \left( 4 a_3^3 a_1 a_5^2-18 a_4^3 a_1^2 a_5+
13 a_3^4 a_1^2-a_3^4 a_5^2-7 a_3^3 a_6^3 +9 a_1^2 a_6^4
+\cdots \right) ,
\label{solgg}
\eeqa
where $D_{1,k}$ is the cofactor for a $(1,k)$ element of the $6\times 6$
matrix $[\pa s_i(a)/\pa a_j]$, $i\in {\cal S}$ and
$j=1, \ldots, 6$ \cite{TeYa2}. In (\ref{solgg}) the explicit expression for
$g_6/g_{12}$ is too long to be presented here, and hence suppressed.
Denoting $y_1=g_9/g_{12},\, y_2=g_8/g_{12},\, y_3=g_6/g_{12} $,
we find that the others are expressed in terms of $y_1,\, y_2$
\beq
 {g_2\over g_{12}}={D_{2,2}\over D_{2,12}}=y_1^2 y_2, \hskip10mm
{g_5\over g_{12}}={D_{2,5}\over D_{2,12}}=y_1 y_2.
\label{solgg2}
\eeq
This means that our superpotential specified with Casimirs (\ref{casimir})
realizes the $SU(2)\times U(1)^5$ vacua only when the coupling constants
are subject to the relation (\ref{solgg2}).

Notice that reading off degrees of $y_1,\, y_2, \, y_3$ from (\ref{solgg}) 
gives $[y_1]=3,\, [y_2]=4,\, [y_3]=6$. 
Thus, if we regard $y_1,\, y_2,\, y_3$ as variables to
describe the $E_6$ singularity, (\ref{solgg}) and (\ref{solgg2}) may be
identified as relevant monomials in versal deformations of the $E_6$
singularity. In fact we now point out an intimate relationship between 
classical solutions corresponding to the symmetry breaking 
$E_6 \supset SU(2)\times U(1)^5$ and the $E_6$ singularity.
For this we examine the superpotential (\ref{treem})
at classical solutions 
\beqa
W_{cl}&=&g_{12}\sum_{k\in {\cal S}} 
\left( {g_k\over g_{12}} \right) s_k^{cl}(a) \CR
&=& g_{12}\, \left( s_2^{cl}y_1^2 y_2+s_5^{cl}y_1y_2+s_6^{cl}y_3
+s_8^{cl} y_2+s_9^{cl} y_1+s_{12}^{cl} \right).
\label{wcl1}
\eeqa
Evaluating the RHS with the use of (\ref{vaccon})-(\ref{solgg2}) leads to
\beq
W_{cl}= -g_{12}\, \left(2 y_1^2 y_3+y_2^3-y_3^2 \right).
\label{wcl2}
\eeq
It is also checked explicitly that
\beqa
 -4 y_1 y_3 &=& 2s_2^{cl}y_1 y_2+s_5^{cl}y_2+s_9^{cl} , \CR
 -3 y_2^2 &=& s_2^{cl}y_1^2+s_5^{cl}y_1+s_8^{cl}, \CR
 -2 y_1^2+2 y_3 &=& s_6^{cl}.
\label{wcl3}
\eeqa

To illustrate the meaning of (\ref{wcl1})-(\ref{wcl3}) let us recall
the standard form of versal deformations of the $E_6$ singularity 
\beq
W_{E_6}(x_1,x_2,x_3;w)=x_1^4+x_2^3+x_3^2+w_2\, x_1^2 x_2+w_5\, x_1x_2
+w_6\, x_1^2+w_8\, x_2+w_9\, x_1+w_{12},
\eeq
where the deformation parameters $w_k$ are related to the $E_6$ Casimirs via
(\ref{casimiro}) \cite{LW}. Then what we have observed in
(\ref{wcl1})-(\ref{wcl3}) is that when we express $w_k$ in terms of $a_i$
as $w_k=w_k^{cl}(a)$ the equations
\beq
W_{E_6}={\pa W_{E_6}\over \pa x_1}={\pa W_{E_6}\over \pa x_2}
={\pa W_{E_6}\over \pa x_3}=0
\eeq
can be solved by
\footnote{We have observed a similar relation between 
the symmetry breaking solutions $SU(r+1)$ (or $SO(2r)$) 
$\supset SU(2)\times U(1)^{r-1}$ and the $A_r$ (or $D_r$) singularity.}
\beq
x_1=y_1(a), \hskip10mm x_2=y_2(a), \hskip10mm
x_3=i \left( y_3(a)-y_1(a)^2-{s_6^{cl}(a) \over 2}\right)
\eeq
under the condition (\ref{vaccon}).
This observation plays a crucial role in our analysis.

When applying the technique of confining phase superpotentials we usually take
all coupling constants $g_k$ as independent moduli parameters. To deal with
$N=1$ $E_6$ theory with fundamental matters, however, we find it 
appropriate to proceed as follows. First of all, motivated by the above
observations for classical solutions, we keep three coupling constants 
$g_6'=g_6/g_{12}$, $g_8'=g_8/g_{12}$ and $g_9'=g_9/g_{12}$ adjustable 
while the rest is fixed as
$g_2'=g_8'g_9'^2,\, g_5'=g_8'g_9'$ with $g_k'=g_k/g_{12}$.
Taking this parametrization it is seen that the equations of motion are
satisfied  by virtue of (\ref{solgg2}) in the $SU(2)\times U(1)^5$ vacua 
(\ref{vaccon}). Note here that originally there exist six classical
moduli $a_i$ among which one is fixed by (\ref{vaccon}) and three are
converted to $g_9'=y_1(a)$, $g_8'=y_2(a)$ and $g_6'=y_3(a)$, 
and hence we are left with two
classical moduli which will be denoted as $\xi_i$. Without loss of generality
one may choose $\xi_2=s_2^{cl}(a)$ and 
$\xi_5=s_5^{cl}(a)$.

We now evaluate the low-energy effective superpotential in the 
$SU(2)\times U(1)^5$ vacua. $U(1)$ photons decouple in the integrating-out 
process. The standard procedure yields the effective superpotential 
for low-energy $SU(2)$ theory \cite{EFGIR},\cite{TeYa2}
\beq
W_L  = -g_{12} \left(2 y_1^2 y_3+y_2^3-y_3^2 \right) \pm 2 \La_{YM}^3,
\label{Wflavor}
\eeq
where the second term takes account of $SU(2)$ gaugino condensation
with $\La_{YM}$ being the dynamical scale for low-energy $SU(2)$ Yang-Mills
theory.
The low-energy scale $\La_{YM}$ is related to the high-energy scale $\La$ 
through the scale matching \cite{TeYa2}
\beqa
\La_{YM}^{6} & =&  g_{12}^2 A(a), \CR
A(a) & \equiv &
\La^{24 - 6 N_f}
\prod_{s=1}^{6}  {\rm det}_{N_f} \left(
\gamma_0+ \gamma_1 (a \cdot \lm_{s}) \right) ,
\label{smr}
\eeqa
where $\lm_{s}$ are weights of ${\bf 27}$ which branch to six $SU(2)$ doublets
respectively under $E_6 \supset SU(2)\times U(1)^5$.
Explicitly they are given in the Dynkin basis as
\beqa
&& \lm_1=(1,\, 0,\, 0,\, 0,\, 0,\, 0),   
\hskip17mm \lm_2=(1,\, -1,\, 0,\, 0,\, 1,\, 0),  \CR
&& \lm_3=(1,\, -1,\, 0,\, 1,\, -1,\, 0), 
\hskip10mm \lm_4=(1,\, -1,\, 1,\, -1,\, 0,\, 0), \CR
&& \lm_5=(1,\, 0,\, -1,\, 0,\, 0,\, 1),  
\hskip14mm \lm_6=(1,\, 0,\, 0,\, 0,\, 0,\, -1).
\eeqa
Notice that $\sum_{s=1}^6\lm_s=3\alpha_1$.

Let us first discuss the $N_f=0$ case, i.e. $E_6$ pure Yang-Mills theory,
for which $A(a)$ in (\ref{smr}) simply equals $\La^{24}$. The vacuum
expectation values are calculated from (\ref{Wflavor})
\beqa
{\pa W_L\over \pa g_{12}}&=&
\langle \widetilde W(y_1,y_2,y_3;s) \rangle 
=-\left(2 y_1^2 y_3+y_2^3-y_3^2\right)
\pm 2 \La^{12}, \CR
{1\over g_{12}}{\pa W_L\over \pa y_1}
&=& \langle {\pa \widetilde W(y_1,y_2,y_3;s)\over \pa y_1} \rangle 
=-4 y_1 y_3, \CR
{1\over g_{12}}{\pa W_L\over \pa y_2}
&=& \langle {\pa \widetilde W(y_1,y_2,y_3;s)\over \pa y_2} \rangle 
=-3 y_2^2, \CR
{1\over g_{12}}{\pa W_L\over \pa y_3}
&=& \langle {\pa \widetilde W(y_1,y_2,y_3;s)\over \pa y_3} \rangle 
=-2 y_1^2+2 y_3,
\label{vevym}
\eeqa
where $y_1, y_2, y_3$ and $g_{12}$ have been treated as independent
parameters as discussed before and
\beq
\widetilde W (y_1,y_2,y_3;s)=s_2\, y_1^2 y_2+s_5\, y_1 y_2
        +s_6\, y_3+s_8\, y_2+s_9\, y_1+s_{12}.
\eeq
Define a manifold by ${\cal W}_0=0$ with four coordinate variables
$z,\, y_1,\, y_2, \, y_3 \in {\bf C}$ and
\beq
{\cal W}_0 \equiv z+ {\La^{24}\over z}
-\left( 2 y_1^2 y_3+y_2^3-y_3^2
+ \widetilde W(y_1,y_2,y_3;s ) \right)=0.
\label{mfd}
\eeq
It is easy to show that the expectation values (\ref{vevym}) parametrize the 
singularities of the manifold where
\beq
{\pa {\cal W}_0\over \pa z}={\pa {\cal W}_0\over \pa y_1}
={\pa {\cal W}_0 \over \pa y_2}={\pa {\cal W}_0\over \pa y_3}=0.
\eeq
Making a change of variables $y_1=x_1,\; y_2=x_2,\; y_3=-ix_3+x_1^2+s_6/2$ 
in (\ref{mfd}) we have
\beq
z+{\La^{24}\over z}- W_{E_6}(x_1,x_2,x_3;  w )=0.
\label{e6ym}
\eeq
Thus the ALE space description of $N=2$
$E_6$ Yang-Mills theory \cite{KLMVW},\cite{LW} is obtained from 
the $N=1$ confining phase superpotential.

We next turn to considering the fundamental matters. In the $N=2$ limit we
have $A(a)=\La^{24-6N_f} \cdot 8^{N_f} \prod_{i=1}^{N_f} f(a, m_i)$ 
with $f(a,m)=\prod_{s=1}^6 (m+a\cdot \lm_s)$. After some algebra we find 
\beq
f(a,m)=m^6+2\xi_2 m^4-8m^3y_1+\left( \xi_2^2-12 y_2 \right) m^2
+4\xi_5 m-4 y_2 \xi_2-8y_3 ,
\eeq
where we have used (\ref{casimir})-(\ref{solgg}).
Recall that, in viewing (\ref{Wflavor}),
we think of $(y_1, y_2, y_3, \xi_2, \xi_5, g_{12})$ as six independent
parameters. Then the quantum expectation values are given by
\beqa
{\pa W_L\over \pa g_{12}}&=&
\langle \widetilde W(y_1,y_2,y_3;s) \rangle 
=-\left( 2 y_1^2 y_3+y_2^3-y_3^2\right)\pm 2 \sqrt{A(y_1,y_2,y_3; \xi ,m)}, \CR
{1\over g_{12}}{\pa W_L\over \pa y_1}
&=& \langle {\pa \widetilde W(y_1,y_2,y_3;s)\over \pa y_1} \rangle
=-4 y_1 y_3 \pm 2 \frac{ \pa }{\pa y_1}\sqrt{A(y_1,y_2,y_3; \xi ,m)}, \CR
{1\over g_{12}}{\pa W_L\over \pa y_2}
&=& \langle {\pa \widetilde W(y_1,y_2,y_3;s)\over \pa y_2} \rangle
=-3 y_2^2 \pm 2 \frac{ \pa }{\pa y_2}\sqrt{A(y_1,y_2,y_3; \xi ,m)}, \CR
{1\over g_{12}}{\pa W_L\over \pa y_3}
&=& \langle {\pa \widetilde W(y_1,y_2,y_3;s)\over \pa y_3} \rangle
=-2 y_1^2+2 y_3 \pm 2 \frac{ \pa }{\pa y_3}\sqrt{A(y_1,y_2,y_3; \xi ,m)}.
\label{vevmat}
\eeqa
Similarly to the $N_f=0$ case one can check that these expectation values 
satisfy the singularity condition for a manifold defined by
\beq
z+ {1 \over z} A(y_1,y_2,y_3;\xi,m)-\left( 2 y_1^2 y_3+y_2^3-y_3^2
+ \widetilde W(y_1,y_2,y_3; s) \right)=0.
\label{ye6ale}
\eeq

Note that $s_k$ in $\widetilde W$ 
are quantum moduli parameters. What about $\xi_2,\, \xi_5$ in 
the one-instanton factor $A$? Classically we have $\xi_i=s_i^{cl}$
as was seen before. The issue is thus whether the classical relations
$\xi_i=s_i^{cl}$ receive any quantum corrections at the singularities.
If there appear no quantum corrections, $\xi_i$ in $A$ can be replaced by
quantum moduli parameters $s_i$. Let us simply assume here that
$\xi_i= s_i^{cl}=\bra s_i\ket$ for $i=2,5$ in the $N=1$ $SU(2)\times U(1)^5$ 
vacua. This assumption seems quite plausible as long as we have inspected 
possible forms of quantum corrections due to gaugino condensates.

Now we find that Seiberg-Witten geometry of $N=2$ supersymmetric QCD 
with gauge group $E_6$ is described by
\beq
z+{1\over z}A(x_1,x_2,x_3; w,m)
- W_{E_6}(x_1,x_2,x_3; w)=0,
\label{e6ale}
\eeq
where a change of variables from $y_i$ to $x_i$ as in (\ref{e6ym}) 
has been made in (\ref{ye6ale}) and 
\beqa
&& A(x_1,x_2,x_3; w,m)  \CR
&=& \La^{24-6N_f} \cdot 8^{N_f} 
\prod_{i=1}^{N_f} \left( {m_i}^6+2 w_2 {m_i}^4-8{m_i}^3x_1 
+\left( w_2^2-12 x_2 \right) {m_i}^2 \right. \CR
&& \left. \hspace{3cm} 
+4 w_5 {m_i}-4 w_2 y_2 -8(x_1^2-ix_3+w_6/2) \right).
\eeqa
The manifold takes the form of ALE space of type $E_6$ fibered over 
the base ${\bf CP}^1$. Note an intricate dependence of the fibering data over
${\bf CP}^1$ on the hypermultiplet masses. This is in contrast with the
ALE space description of $N=2$ $SU(N_c)$ and $SO(2N_c)$ gauge theories with
fundamental matters. 
In (\ref{e6ale}), letting $m_i \rightarrow \infty$ while keeping 
$\La^{24-6N_f} \prod_{i=1}^{N_f} m_i^6 \equiv \La_0^{24}$ finite we
recover the pure Yang-Mills result (\ref{e6ym}).

As a non-trivial check of our proposal (\ref{e6ale}) let us examine
the semi-classical singularities. In the semi-classical limit $\Lambda
\rightarrow 0$ the discriminant $\Delta$ for (\ref{e6ale}) is expected
to take the form $\Delta \propto \Delta_G \Delta_M$ where $\Delta_G$  is a
piece arising from the classical singularities associated with the gauge
symmetry enhancement and $\Delta_M$ represents the semi-classical singularities
at which squarks become massless. When the $N_f$ matter hypermultiplets belong
to the representation $\rep$ of the gauge group $G$ we have
\beq
\Delta_M=\prod_{i=1}^{N_f}{\rm det}_{d \times d}(m_i{\bf 1}-\Phi^{cl})
=\prod_{i=1}^{N_f}P_G^\rep(m_i; u),
\label{delM}
\eeq
where $d={\rm dim}\,\rep$, $m_i$ are the masses, $\Phi^{cl}$ denotes the 
classical Higgs
expectation values and $P_G^\rep(x; u)$ is the characteristic polynomial
for $\rep$ with $u_i$ being Casimirs constructed from $\Phi^{cl}$.

For simplicity, let us consider the case in which all the quarks have equal 
bare masses. Then we can change a variable $x_3$ to $\tilde{x}_3$ so that
$A=A(\tilde{x}_3; w,m)$ is independent of $x_1$ and $x_2$.
Eliminating $x_1$ and $x_2$ from (\ref{e6ale}) by the use of
\beq
{\pa W_{E_6}\over \pa x_1}={\pa W_{E_6}\over \pa x_2}=0,
\label{aaa}
\eeq
we obtain a curve which is singular at the discriminant locus of (\ref{e6ale}).
The curve is implicitly defined through
\beq
\overline{W}_{E_6} \left(\tilde{x}_3; w_i-\delta_{i, 12} 
\left( z+\frac{A \left(\tilde{x}_3; w, m \right)}{z} \right) \right)\
=0,
\eeq
where $\overline{W}_{E_6}(\tilde{x}_3; w_i)
=W_{E_6}(x_1(\tilde{x}_3, w_i),x_2(\tilde{x}_3, w_i),\tilde{x}_3; w_i)$ and
$x_1(\tilde{x}_3, w_i)$, $x_2(\tilde{x}_3, w_i)$ are solutions
of (\ref{aaa}). Now the values of $\tilde{x}_3$ and $z$ at singularities 
of this
curve can be expanded in powers of $\La^{\frac{24-6 N_f}{2}}$. Then it is more
or less clear that the classical singularities corresponding to massless
gauge bosons are produced. Furthermore, if we denote as $R(\overline{W},A)$ 
the resultant of $\overline{W}_{E_6}(\tilde{x}_3; w_i)$
and $A \left(\tilde{x}_3;w, m \right)$, then $R(\overline{W},A)=0$ yields
another singularity condition of the curve in the limit 
$\Lambda \rightarrow 0$.
We expect that $R(\overline{W},A)=0$ corresponds to the semi-classical
massless squark singularities as is observed in the case of $N=2$ $SU(N_c)$
QCD \cite{HO},\cite{GPR}.
Indeed, we have checked this by explicitly computing 
$R(\overline{W},A)$ at sufficiently many points in the
moduli space. For instance, taking $w_2=2,w_5=5,w_6=7,w_8=9,w_9=11$ and 
$w_{12}=13$ in the $N_f=1$ case,  we get
\beqa
&& R(\overline{W},A) \CR
& = & m^2 \left( 3\,m^{10}+12\,m^{8} +\cdots \right)
\left( 26973 m^{27}+258552\,m^{25}+\cdots \right)^3 \CR
& & \left( m^{27} + 24\,m^{25} + 240\,m^{23} + 240\,m^{22} 
+ 2016\,m^{21} + 3360\,m^{20} + 16416\,m^{19} \right.  \CR
 & & + 34944\,m^{18}+ 88080\,m^{17} + 216576\,m^{16} + 448864\,m^{15} +
607488\,m^{14} \CR
 & &  + 2198272\,m^{13}- 296000\,m^{12}+ 4177792\,m^{11}
- 3407104\,m^{10} + 7796224\,m^{9} \CR
 & &  + 10664448\,m^{8}- 31708160\,m^{7}+ 41183232\,m^{6}
- 21889792\,m^{5} + 15575040\,m^{4} \CR
 & &  \left. - 17125120\,m^{3}- 38456320\,m^{2}- 3461120\,m + 9798656 \right) ,
\label{bbb}
\eeqa
while the $E_6$ characteristic polynomial for ${\bf 27}$ is given by
\beqa
&& P_{E_6}^{\bf 27}(x; u)  \CR
&=& x^{27}+12 w_2 x^{25}+60 w_2^2 x^{23}+48 w_5 x^{22}
+\left( 96w_6+168w_2^3 \right) x^{21}+336w_2 w_5 x^{20} \CR
&& +\left( 528w_2w_6+294w_2^4+480w_8 \right) x^{19}
+\left( 1344w_9+1008w_2^2w_5 \right) x^{18}+\cdots .
\eeqa
We now find a remarkable result that the last factor of (\ref{bbb})
precisely coincides with $P_{E_6}^{\bf 27}(m; u)$! Hence the manifold described
by (\ref{e6ale}) correctly produces all the semi-classical singularities
in the moduli space of $N=2$ supersymmetric $E_6$ QCD.

If we choose another form of the superpotential (\ref{treem}), say, the 
superpotential with $s_i=w_i$ for $i \in {\cal S}$ instead of (\ref{casimir})
we are unable to obtain $\Delta_M$ in (\ref{delM}). As long as we have
checked the choice made in (\ref{casimir}) is judicious in order to pass the
semi-classical test. At present, we have no definite recipe to fix the
tree-level superpotential which produces the correct semi-classical 
singularities, though it is possible to proceed by trial and error. In fact
we can find Seiberg-Witten geometry for $N=2$ $SO(2 N_c)$ gauge theory with
spinor matters and $N=2$ $SU(N_c)$ gauge theory with antisymmetric
matters \cite{TeYa3}.

In our result (\ref{e6ale}) it may be worth mentioning that 
the gaussian variable $x_3$
of the $E_6$ singularity appears in the fibering term. An analogous structure 
is observed for $N=2$ $SO(10)$ gauge theory with spinors and vectors
in \cite{AG}. Their result reads
\beq
z+{\Lambda^{2b}\over z}B(x_1,x_3; v)-W_{D_5}(x_1,x_2,x_3; v)=0,
\eeq
where $b=8-(6-n)-2(4-n)=3n-6$ is the coefficient of the one-loop beta function
and 
\beqa
&& W_{D_5}(x_1,x_2,x_3; v)
=x_1^4+x_1x_2^2-x_3^2+v_2x_1^3+v_4x_1^2+v_6x_1+v_8+v_5x_2, \CR
&& B(x_1,x_3; v)=x_1^{6-n}
\left( x_3-{1\over 8}\left( 4v_4-v_2^2+4v_2x_1+8x_1^2 \right) \right)^{4-n}
\eeqa
for $SO(10)$ with massless $(6-n)$ fundamentals and $(4-n)$ spinors. Here $v_i$
stand for the $SO(10)$ Casimirs. It is tempting to suspect that the above $E_6$
and $SO(10)$ results are related through the Higgs mechanism under the
symmetry breaking $E_6 \supset SO(10) \times U(1)$.

There is no difficulty in using our method to find Seiberg-Witten geometry
in the form of ALE fibrations for $N=2$ QCD with $E_7$ gauge group although 
more computer powers are obviously required. Finally, to compare the present 
results, it is desired to work out
exceptional Seiberg-Witten geometry with fundamental matters in the
framework of Calabi-Yau geometric engineering \cite{KKV}-\cite{KMV}.

\vskip10mm

The work of S.T. is supported by JSPS Research Fellowship for Young 
Scientists. The work of S.K.Y. was supported in part by the
Grant-in-Aid for Scientific Research on Priority Area 
``Physics of CP violation'', the Ministry of Education, Science and Culture, 
Japan.

\newpage


\end{document}